\documentclass[twocolumn,twoside,slac_two]{revtex4}
\usepackage{graphicx}
\usepackage{fancyhdr}
\usepackage{relsize}
\newcommand \babar{\mbox{\slshape B\kern-0.1em{\smaller A}\kern-0.1em
    B\kern-0.1em{\smaller A\kern-0.2em R}}}

\begin{document}

\title{Inclusive and Exclusive $|V_{ub}|$}

\author{A. Petrella on behalf of \babar\ and Belle Collaborations}
\affiliation{Universit\`a degli Studi and INFN, Ferrara, Italy}

\begin{abstract}
  The current status of the determinations of CKM matrix element
  $|V_{ub}|$ via exclusive and inclusive charmless semileptonic $B$ decays
  is reviewed.
\end{abstract}

\maketitle

\thispagestyle{fancy}

\section{Introduction}
In the Standard Model (SM), weak transitions between quark
flavours are described by the elements $V_{ij}$ of the Cabibbo
Kobayashi Maskawa (CKM)\cite{CKM}
matrix. Theory does not predict the magnitude of the elements which
therefore must be determined experimentally. The matrix is unitary
by construction and one of the unitarity conditions,
$V_{ud}V^*_{ub}+V_{cd}V_{cb}^*+V_{td}V_{tb}^*=0$, can be
geometrically represented as a triangle in the complex plane
$(\bar\rho,\bar\eta)$\footnote{$\bar{\rho}=(1-\lambda^2/2)\rho$,
  $\bar{\eta}=(1-\lambda^2/2)\eta$, where $\lambda=V_{us}$ and
  $A\lambda^3(\rho-i\eta)=V_{ub}$. $\rho$, $\eta$ and $A$ are defined
  in the Wolfenstein's CKM parametrization~\cite{WOLF}.}, the well
known Unitarity Triangle (UT). Any non zero value of the $\bar\eta$
parameter is an indication of $CP$ violation. The angles and sides of
the UT can be measured by studying $B$ meson decays.

The Unitarity Triangle analysis shows an impressive success of the
CKM picture in describing $CP$ violation in the SM, but, as the
experimental results become increasingly precise, a slight
disagreement between the angle $\beta$, characterising indirect $CP$
violation in $b\rightarrow\bar{c}cs$ transitions and currently known at
the 4\% level, and $|V_{ub}|/|V_{cb}|$ has appeared in the UT fit. This
disagreement could be due to some problems with theoretical
calculations impacting on $|V_{ub}|$ determinations. Tree-level
processes are essentially immune to contributions from new physics, so
studying semileptonic $B$ decays and therefore determining $|V_{cb}|$ and
$|V_{ub}|$ is a way to test the electroweak sector of the SM. While
the determination of $|V_{cb}|$ is at the 2\% level~\cite{BUCH}, the
uncertainty on $|V_{ub}|$ is still at the 8\% level. The need for an
improvement in the precision on $|V_{ub}|$ is therefore evident.

In the following we will present the current status and outlook
regarding experimental determinations of $|V_{ub}|$.

\section{Semileptonic $B$ Decays}
The theoretical description of charmless semileptonic $B$ decays is at
a mature stage. $\bar{B}\rightarrow X_u\ell\bar\nu$ decays provide the
cleanest way to measure $|V_{ub}|$ since the leptonic and hadronic
part of the weak current factorize into two terms not interacting
between each other, resulting in an easy theoretical description at
the parton level, even though uncertainties arise when introducing QCD
calculations to describe the hadronization process. Given the fact
that the $b$ quark mass is considerably larger than the scale
$\Lambda_{QCD}$ that determines the low energy hadron physics, the
total rate can be expanded in powers of $\Lambda_{QCD}/m_b$ and
$\alpha_S$, separating non-perturbative and perturbative physics. 

Two main experimental approaches are used to measure $|V_{ub}|$ from
$\bar{B}\rightarrow X_u\ell\bar\nu$ decays, depending on the choice
being made between integrating over all possible charmless final
states or selecting a particular one: inclusive and exclusive. The
first approach provides higher signal efficiency while the second
gives a better background rejection. Theoretical inputs are needed by
both approaches to model the hadronization, but since they rely on
independent calculations, they provide two complementary
determinations of $|V_{ub}|$.

\section{Experimental Techniques}
The most recent measurements of charmless semileptonic $B$ decays have
been performed by the \babar, Belle and CLEO experiments. These
experiments record $e^+e^-$ collisions at the energy of the
$\Upsilon(4S)$, a $b\bar{b}$ bound state that decays predominantly to
$B^0\bar{B}^0$ or $B^+B^-$ mesons. The main backgrounds for
$b\rightarrow u\ell\bar\nu$ transitions are the more abundant
$b\rightarrow c\ell\bar\nu$
(rate $\sim50$ times larger), the continuum background coming from
$e^+e^-\rightarrow q\bar{q},\ q = (u,d,s,c)$ and, where applicable,
combinatorial background due to random association of tracks in the
reconstruction of a $B$ meson. There are three established
experimental techniques employed to select signal events, that differ
on the reconstruction of the second $B$ in the event ({\it tag} side),
and are described below.

\subsection{Untagged Method}
In the untagged method the $B$ recoiling against the signal $B$ is not
explicitly reconstructed. With this technique, the neutrino
four-momentum is inferred from the difference between the
four-momentum of the colliding beam particles and the sum of the
four-momenta of all the charged and neutral particles detected in a
single event. The kinematic consistency of a $B_{tag}$ candidate with
a $B$ meson decay is evaluated using two variables: the beam-energy
substituted mass $m_{ES} \equiv \sqrt{s/4-|p^*_B|^2}$, and the energy
difference $\Delta E \equiv E^*_B -\sqrt{s}/2$. Here $ \sqrt{s}$ is
the total CM energy, and $p^*_B$ and $E^*_B$ denote the magnitude of
the 3-momentum and energy of the $B_{tag}$ candidate in the CM
frame. For correctly identified $B_{tag}$ decays, the $m_{ES}$
distribution peaks at the $B$ meson mass, while $\Delta E$ is
consistent with zero.

The untagged method offers higher signal efficiency ($\sim5\%$) with
respect to the other two methods but due to the poor resolution on the
neutrino 4-momentum has lower purity.

\subsection{Semileptonic Tag Method}
In the semileptonic method, a $B\rightarrow \bar{D}^{(*)}\ell\nu$ decay is
reconstructed in the tag side. Several $\bar{D}$ and $\bar{D}^*$ decay
modes are used for tagging. The presence of two neutrinos requires other
kinematical constraints in order to separate signal events from
backgrounds. With respect to the untagged method, the semileptonic tag
provides lower efficiency ($\sim 1\%$) but higher purity.

\subsection{Hadronic Tag Method}
In this method the tag side is reconstructed as a decay of the type
$B\rightarrow \bar{D}^{(*)}Y$, where $Y$ represents a linear
combination of charged and neutral pions and kaons. Several decay
combinations are taken into account. $\Delta E$ and $m_{ES}$ variables
are used to check the consistency of the reconstructed $B$
candidate. Since the $B_{tag}$ is fully reconstructed, the kinematics
of the event is completely constrained and charge and flavour of the
signal $B$ can be inferred. Given the good neutrino 4-momentum
resolution provided by this method, other kinematical variables, such
as the leptonic squared invariant mass $q^2$, the missing mass squared
$mm^2$ or the hadronic invariant mass $m_X$ can be exploited to
separate the background from the $b\rightarrow u$ signal events. The
fallback of this method is the very low tag efficiency (at the order
of $10^{-3}$).

\section{Exclusive $|V_{ub}|$ Determinations}
In the exclusive approach, the measured branching fraction for a
specific charmless decay channel,
e.g. $\bar{B}\rightarrow\pi\ell\bar\nu$, is converted into $|V_{ub}|$
using theoretical calculations of the form factors (FF) which
parametrize QCD effects. In particular, for
$\bar{B}\rightarrow\pi\ell\bar\nu$ decays, the differential branching
fraction as function of $q^2$ is proportional to $|V_{ub}||f_+(q^2)|$,
the latter term of the product being the FF. Experiments measure
$|V_{ub}||f_+(q^2)|$ and information on the shape and normalization of
$f_+(q^2)$ must come from theory. Several FF calculations are
available, based on quark models~\cite{ISGW}, lattice
QCD~\cite{HPQCD,FNAL} and Light Cone Sum Rules
(LCSR)~\cite{LCSR}. Lattice QCD and LCSR calculations have validity in
complementary $q^2$ ranges, giving predictions for
$q^2>16$~GeV$^2$/$c^4$ and $q^2<14$~GeV$^2$/$c^4$ respectively.

\begin{figure}[ht]
  \centering
  \includegraphics[width=70mm]{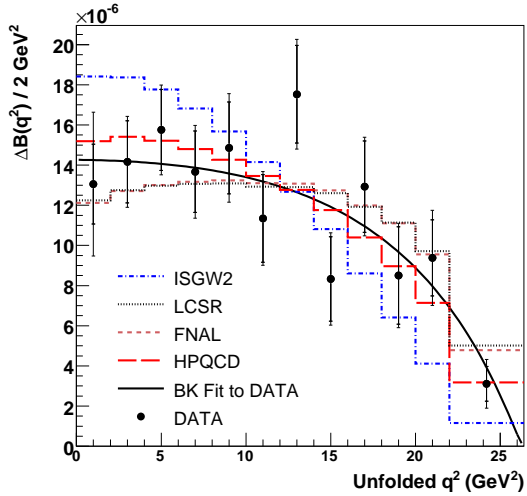}
  \caption{$\Delta\mathcal{B}(B^0\rightarrow\pi^-\ell^+\nu)$ as function of
    $q^2$ measured by \babar~\cite{unt}. The solid black curve shows the
    result of the fit of the BK parametrization~\cite{BK} to the
    data. Other FF calculations~\cite{HPQCD,FNAL,LCSR,ISGW} are
    also compared to data.} \label{FF_figure}
\end{figure}

With more and more statistics provided by the $B$-Factories, it has become
possible to measure branching ratios in different $q^2$ intervals and
compare the predicted FF shapes to experimental
data. Figure~\ref{FF_figure} shows the differential partial branching
ratio spectrum as function of $q^2$ for $B^0\rightarrow\pi^-\ell^+\nu$
decays measured with an untagged analysis performed
by~\babar~\cite{unt}. This analysis has shown that FF calculations
based on quark models are not consistent with data distributions.

Table~\ref{excl_table} lists a summary of published branching ratio
determinations for $B\rightarrow\pi\ell^+\nu$ decays. All the
measurements are consistent within the experimental
uncertainties. Among all the methods, the untagged one provides the
most precise measurement, having an uncertainty of approximately 7\% in
the branching ratio determination.
\begin{table}[h]
  \begin{center}
    \caption{$\mathcal{B}(B\rightarrow\pi\ell^+\nu)$ measurements for
      different experiments and tagging techniques. U, SL and Had indicate
      untagged, semileptonic tag and hadronic tag methods respectively. Errors
      on branching ratios are statistical and systematic.}
    \begin{tabular}{|c|c|c|l|}
      \hline 
      \textbf{Mode} & \textbf{$B\bar{B}$} & \textbf{Branching
        Ratio} & \textbf{Exp./Tag}\\
      & $[10^6]$ & $[10^{-4}]$ & \\
      \hline
      $B^0\rightarrow\pi^-\ell^+\nu$ & 227 & $1.46\pm0.07\pm0.08$ &
      \babar~\cite{unt} U\\
      \cline{2-4}
      & 15.4 & $1.37\pm0.15\pm0.11$ & CLEO~\cite{CLEOunt} U\\
      \cline{2-4}
      & 232 & $1.12\pm0.25\pm0.10$ & \babar~\cite{babarsl} SL\\
      \cline{2-4}
      & 275 & $1.38\pm0.19\pm0.14$ & Belle~\cite{bellesl} SL\\
      \cline{2-4}
      & 232 & $1.07\pm0.27\pm0.15$ & \babar~\cite{babarsl} Had\\
      \hline
      $B^+\rightarrow\pi^0\ell^+\nu$ & 232 & $0.73\pm0.18\pm0.08$ &
      \babar~\cite{babarsl} SL\\
      \cline{2-4}
      & 275 & $0.77\pm0.14\pm0.08$ & Belle~\cite{bellesl} SL\\
      \cline{2-4}
      & 232 & $0.82\pm0.22\pm0.11$ & \babar~\cite{babarsl} Had\\
      \hline
    \end{tabular}
    \label{excl_table}
  \end{center}
\end{table}
The world average computed by the HFAG~\cite{HFAG} group is
$\mathcal{B}(B^0\rightarrow\pi^-\ell^+\nu)=(1.38\pm0.06\pm0.07)\times10^{-4}$,
where the first error is statistical and the second due to systematic
uncertainties.

\begin{figure}[ht]
  \centering
  \includegraphics[width=80mm]{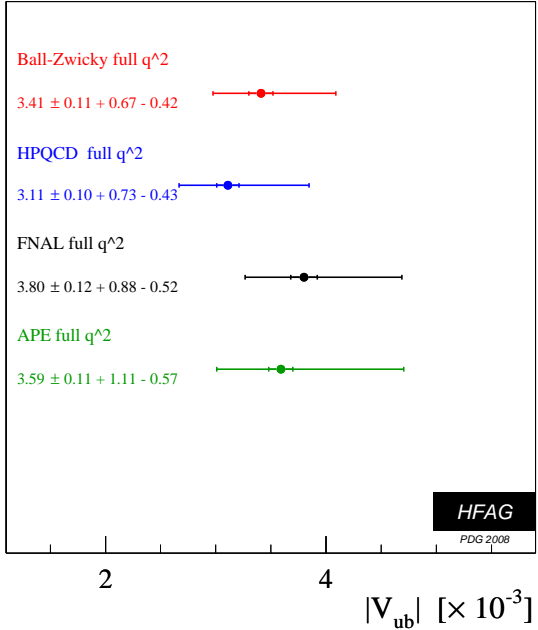}
  \caption{Comparison of exclusive $|V_{ub}|$ determinations for
    different form factor calculations for the full $q^2$
    range.} \label{VubExcl_figure}
\end{figure}

According to the various FF calculations, different $|V_{ub}|$ values
can be computed and are shown in Fig.~\ref{VubExcl_figure}, for the
full $q^2$ range. With the exclusive approach, the central value for
$|V_{ub}|$ lies in the interval $[3.11,3.80]\times10^{-3}$, in good
agreement with the indirect determination of $|V_{ub}|$ performed by
UT fit collaboration:
$|V_{ub}|_{UTfit}=(3.44\pm0.16)\times10^{-3}$~\cite{UTFIT}. The exclusive
determinations are still limited by the theoretical uncertainties on
the knowledge of the FF, which contribute up to 23\% to the total
error. 

Branching ratio measurements of a $B$ meson decaying into other
charmless semileptonic final state ($\rho$, $\eta/\eta'$, $\omega$)
have been performed (world averages can be found on HFAG
website~\cite{HFAG}); however theory calculations necessary to convert
these measurements into $|V_{ub}|$ values are not yet mature.

\section{Inclusive $|V_{ub}|$ Determinations}
The full rate of inclusive charmless $B$ decays is computed within the
Operator Product Expansion (OPE) with a theory uncertainty
of $\sim 5\%$, mainly due to uncertainty on the $b$ quark mass. In
practice the accessible rate is reduced since it is necessary to
exploit kinematical variables that describe the semileptonic decays in
order to suppress the overwhelming background from $b\rightarrow c$
transitions; this restricts the measurement to phase space regions
where particles containing charm cannot be produced. The drawback of
this approach is on the theory side since calculating partial widths
in regions of phase space where $\bar{B}\rightarrow X_c\ell\bar\nu$
are suppressed is very challenging, as the HQE convergence in these
regions is spoiled and a non-perturbative distribution function,
the shape function (SF)\cite{SF1,SF2}, whose form is unknown, needs to
be introduced. Weak annihilation and other non-perturbative effects
need to be modeled too. 

The shape function is a universal property of $B$ mesons at leading
order, however sub-leading shape functions arise at each order in the
$1/m_b$ expansion. SF parameters can be constrained by measuring
moments of inclusive distributions from $\bar{B}\rightarrow
X_c\ell\bar\nu$ and $\bar{B}\rightarrow X_s\gamma$ decays which are related
to the same heavy quark parameters ($m_b$, the $b$ quark mass and
$\mu_\pi^2$, the square of the kinetic energy of the $b$ quark in the
$B$ meson).

In recent years, many theoretical calculations have become
available, either based on the OPE approach~\cite{BLNP,LNR,BLL,GGOU}
or on models of non-perturbative QCD~\cite{DGE,AC}.

Several kinematical variables are used to separate signal from
$b\rightarrow c$ background, each having its own advantage. The
lepton energy $E_\ell$ is the simplest to measure but the cut applied
to reduce the charmed background restricts the total accessible signal
rate to $\sim10\%$ of the total; moreover the dependence on leading
and subleading SF and weak annihilation corrections may be
substantial. The squared leptonic invariant mass $q^2$ is weakly
sensitive to SF effects, has higher accessible $b\rightarrow u$
fraction, $\sim 20\%$, but is sensitive to weak annihilation
corrections. Much higher signal rate is provided by the hadronic
invariant mass $m_X$ and the light cone momentum
$P_+=E_X-|\vec{p}_X|$, $\sim80\%$ and $\sim70\%$ respectively, but
both depend on SF and subleading SF corrections.
\begin{figure*}[t]
  \centering
  \includegraphics[totalheight=8.5cm]{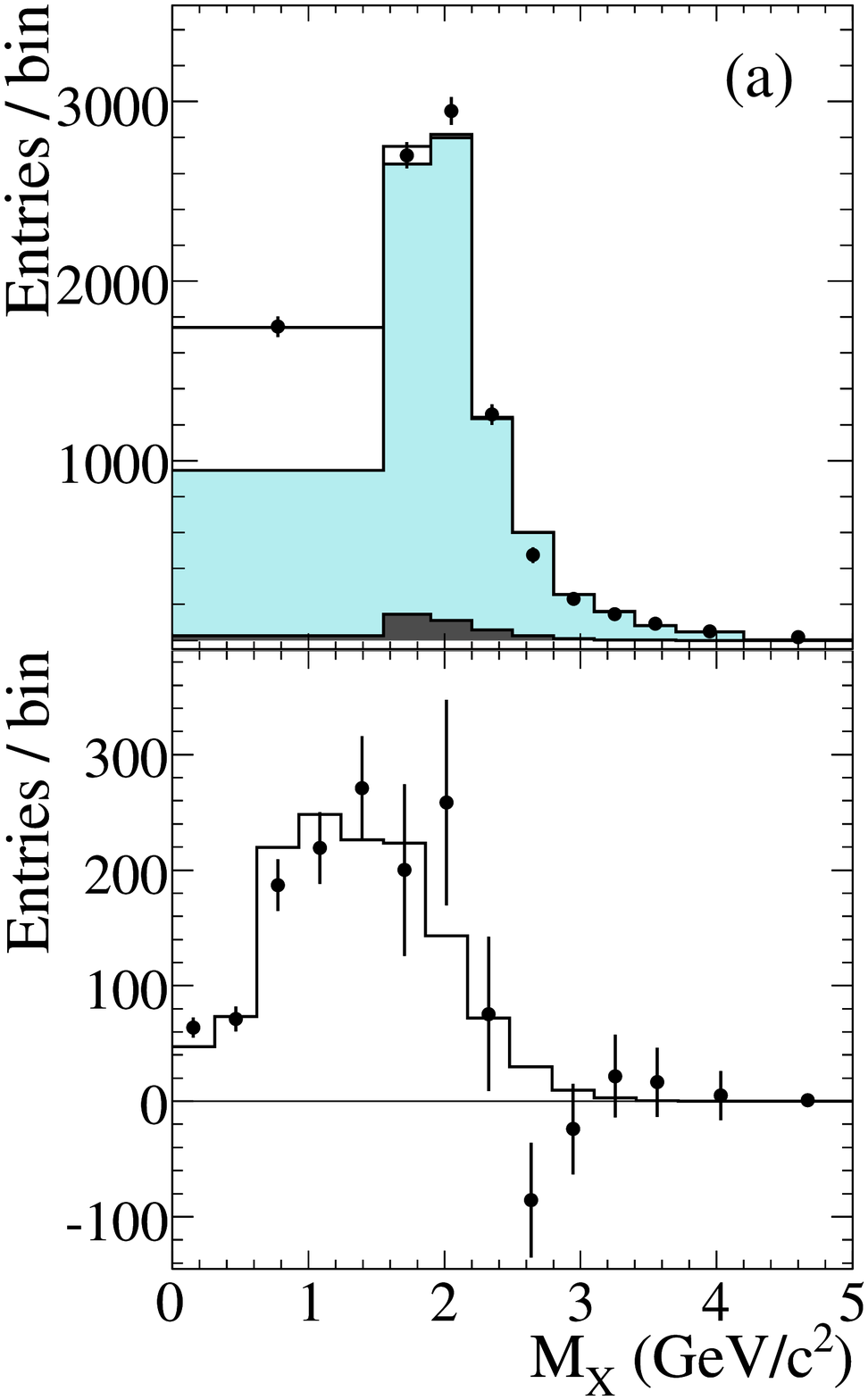}
  \includegraphics[totalheight=8.5cm]{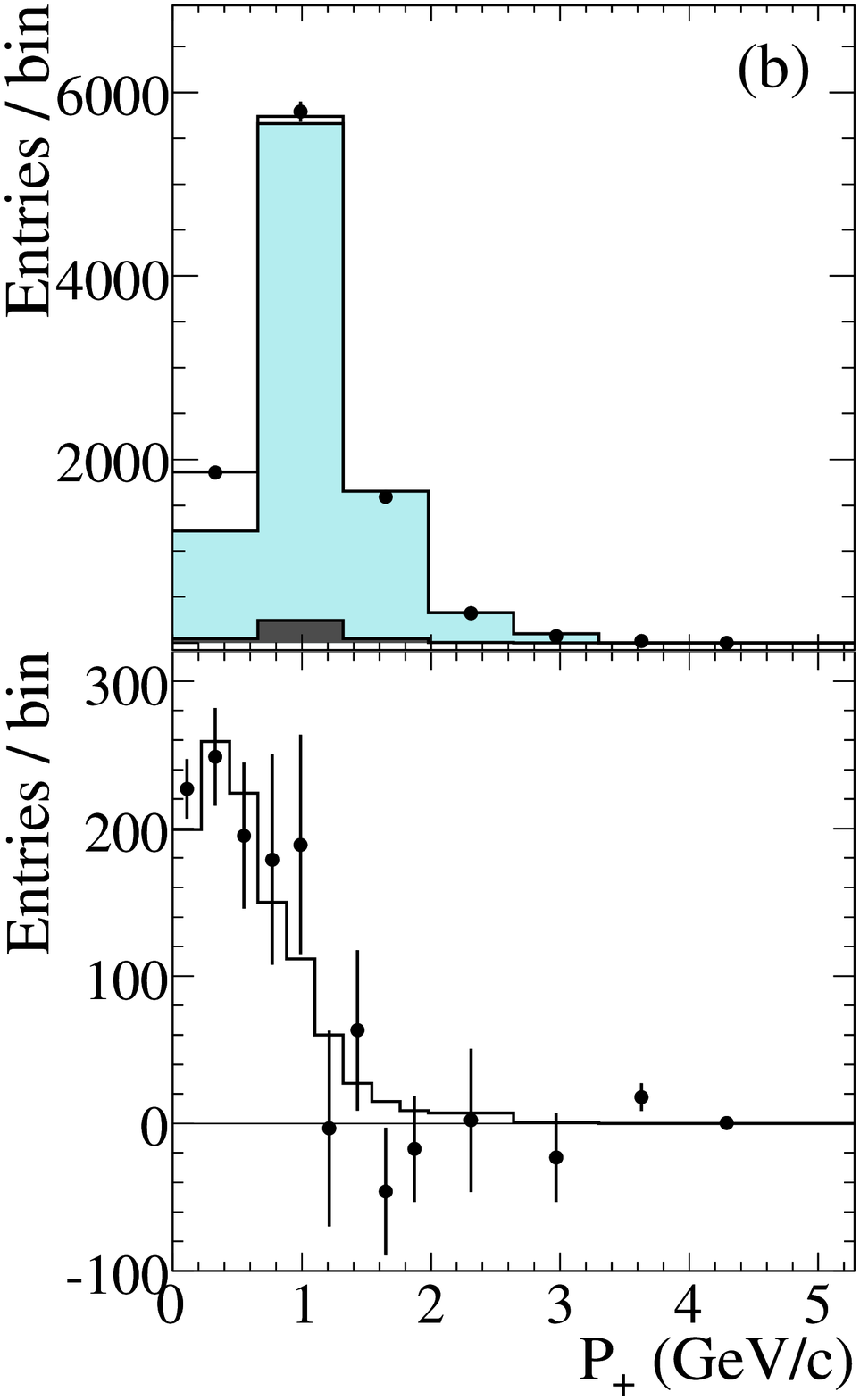}
  \includegraphics[totalheight=8.5cm]{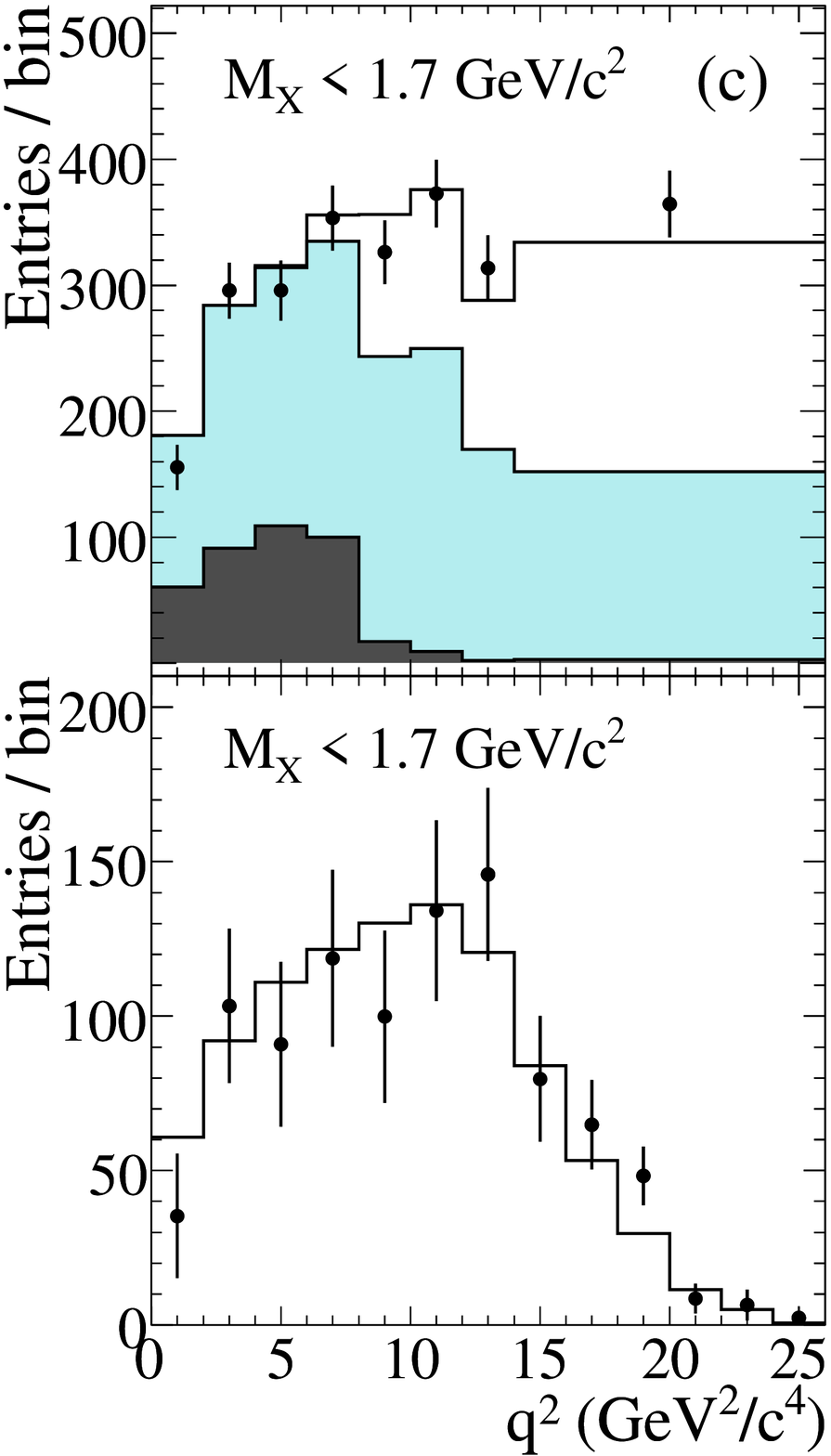}
  \caption{Upper row: $m_X$ (a), $P_+$ (b) and $q^2$ with $m_X<1.7$
    GeV/$c^2$ (c) measured spectra (data points)~\cite{babarhad}. The
    result of the fit to the sum of three Monte Carlo contributions is
    shown in the histograms: $\bar{B}\rightarrow X_u\ell\bar\nu$
    decays generated inside (white) and outside (gray) the signal
    region, and $\bar{B}\rightarrow X_c\ell\bar\nu$ and other
    background (blue). Lower row: corresponding spectra
    for $\bar{B}\rightarrow X_u\ell\bar\nu$ after $\bar{B}\rightarrow
    X_c\ell\bar\nu$ and other background subtraction, rebinned to show
    the shape of the kinematical variables. } \label{babarHad_figure}
\end{figure*}
The most recent inclusive $|V_{ub}|$ determinations have been
performed by the \babar\ experiment using the hadronig tag
technique~\cite{babarhad}. In this analysis, inclusive $m_X$, $P_+$
and $(m_X,q^2)$ distributions have been reconstructed for semileptonic
$B$ decays and measurements of charmless partial branching fractions
have been performed in regions of phase space where the $b\rightarrow
c$ transitions are highly suppressed. Continuum and combinatoric
backgrounds have been subtracted with fits to $m_{ES}$
distributions. Figure~\ref{babarHad_figure} shows the fits of Monte
Carlo $b\rightarrow c$ (blue), $b\rightarrow u$ (white) and other
background (gray) shapes to the measured data (points) from which
partial branching fractions for the signal enhanced region have been
determined. $|V_{ub}|$ values have then been calculated using the
relation
\begin{equation}
  |V_{ub}|=\sqrt{\frac{\Delta\mathcal{B}(\bar{B}\rightarrow
      X_u\ell\bar\nu)}{\tau_B\tilde{\Gamma}_{thy}}}\label{vub_formula}
\end{equation}
where $\tau_B$ is the $B$ lifetime and $\tilde{\Gamma}_{thy}$ are the
theoretical acceptances provided by the cited models. The $|V_{ub}|$
determinations obtained are in good agreement with the ones provided
in a similar analysis by Belle~\cite{Bellehad}, and show
that the measurements based on $m_X$ and $(m_X,q^2)$ are compatible
with theory calculations, while there is a hint that results based on
$P_+$ are somewhat lower than theory predictions and closer to
$|V_{ub}|$ determinations which use exclusive charmless semileptonic
decays.
 
\begin{figure}[ht]
  \centering
  \includegraphics[width=80mm]{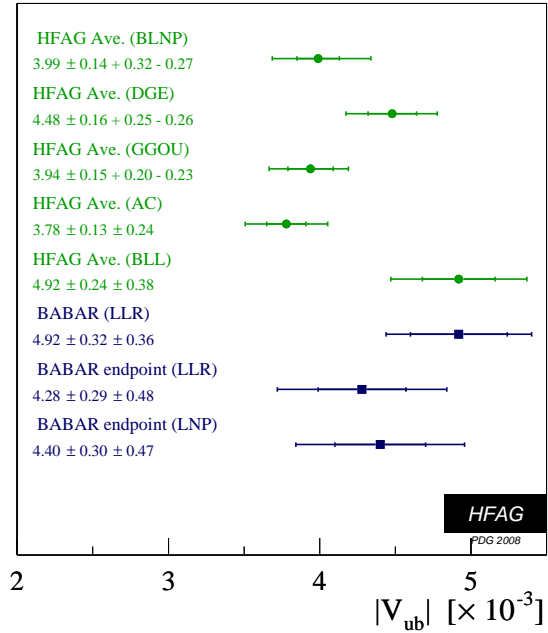}
  \caption{Comparison of inclusive $|V_{ub}|$ values obtained using different
    theoretical calculations.}\label{inclVub_figure}
\end{figure}

HFAG~\cite{HFAG} provides world averages of  $|V_{ub}|$ values
obtained within the
currently available theoretical frameworks and these are listed in
Fig.~\ref{inclVub_figure}. Inclusive charmless semileptonic decays
give $|V_{ub}|$ determinations that are compatible with exclusive
ones, even though with higher values. As is the case for the exclusive
measurements, the dominant uncertainty is due to theory ($\sim7\%$). 

\section{Weak Annihilation}
{\it Weak annihilation} denotes a $B^+\rightarrow X_u\ell^+\nu$ decay in which
the $\bar{b}$ and the spectator $u$ quark forming the $B^+$ meson
annihilate into a $W^+$ boson, and a soft gluon emitted in the
interaction materializes into a charmless final state. The
contribution to the total charmless semileptonic rate is expected to
be small, of the order of 3\%, but can be
relevant when selecting large $q^2$ regions. Weak
annihilation can be experimentally observed as a difference in
the partial decay rates of $B^0\rightarrow X_u^-\ell^+\nu$ and
$B^+\rightarrow X^0_u\ell^+\nu$ at high $q^2$ since it occurs only for
charged $B$ mesons. Measurements performed by \babar~\cite{babarWA}
and CLEO~\cite{CLEOWA} have provided no evidence of weak annihilation
so far, placing the upper limit: $\Gamma_{WA}/\Gamma_{b\rightarrow
  u\ell\nu}<8\%$ at $90\%$ CL.

\section{Conclusion}
The large datasets collected at the $B$-Factories, and the increased
precision of theoretical calculations have allowed an improvement in the
determination of $|V_{ub}|$. However, there are still significant
uncertainties. In the exclusive approach, the most precise measurement
of the pion channel branching ratio is obtained by an untagged
analysis. This very good precision can be reached by tagged analyses
with more data. The problem with exclusive decays is that the strong
hadron dynamics can not be calculated from first principles and the
determination of the form factor has to rely on light-cone sum rules or
lattice QCD calculations. The current data samples allow a comparison
of different FF models with data distributions. With
further developments on lattice calculations, the theoretical error
should shrink to reach the experimental one.

The inclusive approach still provides the most precise $|V_{ub}|$
determinations. With new theoretical calculations, the mild ($2.5\sigma$)
discrepancy with respect to the $|V_{ub}|$ value determined from the
global UT fit has been reduced. As in the exclusive approach,
theoretical uncertainties represent the limiting factor to the precision
of the measurement. Reducing the theoretical uncertainties
to a level comparable with the statistical error is challenging. New
measurements in semileptonic decays of charm mesons could increase the
confidence in theoretical calculations and related uncertainties.

\end{document}